\documentclass[runningheads]{llncs}
\usepackage[T1]{fontenc}
\usepackage[table,xcdraw]{xcolor}
\usepackage{graphicx}
\usepackage{balance}
\usepackage{soul}
\usepackage{booktabs}
\usepackage{multirow}
\usepackage{float}
\usepackage{url}
\usepackage{tcolorbox}
\usepackage{tabularx} 
\usepackage[noadjust]{cite}
\usepackage[hidelinks,bookmarks=true,pagebackref=true]{hyperref}
\usepackage{subcaption}
\usepackage{colortbl}
\usepackage{wrapfig}
\usepackage{wrapfig,lipsum,booktabs}

\newcommand{\snippetResults}[1]{
    \begin{figure}[H]
    \vspace{-15pt}
    \setlength{\abovecaptionskip}{0pt} 
    \setlength{\belowcaptionskip}{0pt} 
    \begin{subfigure}{0.32\linewidth}
        \includegraphics[width=\linewidth]{img/snippets_measures/snippet_#1_Average_Heart_Rate.png}
        \caption{Average Heart Rate}
        \label{fig:avgheartrate_#1}
    \end{subfigure}
    \begin{subfigure}{0.32\linewidth}
        \includegraphics[width=\linewidth]{img/snippets_measures/snippet_#1_Time_Spent.png}
        \caption{Time Spent on Snippet}
        \label{fig:timespent_#1}
    \end{subfigure}
    \begin{subfigure}{0.32\linewidth}
        \includegraphics[width=\linewidth]{img/snippets_measures/snippet_#1_correctness.png}
        \caption{\% Correctly Answered}
        \label{fig:correctlyanswered_#1}
    \end{subfigure}
    \caption{Biometrics and Other Measures Averaged per IPC Group for Snippet #1}
    \medskip
    \tiny
    Average heart rate and time spent was further averaged within each IPC grouping (Few, Moderate, Frequent, and Intense IPC). The percentage of participants who scored correctly in each IPC grouping is displayed in the last figure.
    \label{fig:biometrics_other_per_snippet_#1}
    \vspace{-15pt}
\end{figure}
}

\usepackage[utf8]{inputenc}
\usepackage{newunicodechar,graphicx}
\DeclareRobustCommand{\okina}{%
  \raisebox{\dimexpr\fontcharht\font`A-\height}{%
    \scalebox{0.8}{`}%
  }%
}
\newunicodechar{ʻ}{\okina}

\begin{document}
\title{Impostor Syndrome in Final Year Computer Science Students: An Eye Tracking and Biometrics Study}
\titlerunning{Impostor Syndrome in Final Year Computer Science Students}
%
\author{Alyssia Chen\orcidID{0009-0002-5684-8601} \and
Carol Wong \and
Katy Tarrit \and
Anthony Peruma\orcidID{0000-0003-2585-657X}}
\authorrunning{A. Chen et al.}

\institute{University of Hawaiʻi at Mānoa \\
              Honolulu, Hawaiʻi, USA \\
              \email{abc8@hawaii.edu, carolw8@hawaii.edu, katytm7@hawaii.edu, peruma@hawaii.edu}
}

\maketitle              
\begin{abstract}
Imposter syndrome is a psychological phenomenon that affects individuals who doubt their skills and abilities, despite possessing the necessary competencies. This can lead to a lack of confidence and poor performance. While research has explored the impacts of imposter syndrome on students and professionals in various fields, there is limited knowledge on how it affects code comprehension in software engineering. In this exploratory study, we investigate the prevalence of imposter syndrome among final-year undergraduate computer science students and its effects on their code comprehension cognition using an eye tracker and heart rate monitor. Key findings demonstrate that students identifying as male exhibit lower imposter syndrome levels when analyzing code, and higher imposter syndrome is associated with increased time reviewing a code snippet and a lower likelihood of solving it correctly. This study provides initial data on this topic and establishes a foundation for further research to support student academic success and improve developer productivity and mental well-being.

\keywords{Cognitive Load and Performance \and Eye Tracking \and Biometrics \and Heart Rate \and Impostor Syndrome \and Code Comprehension  \and Program Comprehension \and Computer Science \and Undergraduate Students}
\end{abstract}
\section{Introduction}
\label{Section:intro}
Impostor syndrome, also known as impostor phenomenon, fraud syndrome, perceived fraudulence, or impostor experience, is a psychological phenomenon characterized by persistent self-doubt of intellect, skills, or accomplishments and feelings of fraudulence or inadequacy despite evidence of one's competence and accomplishments \cite{Clance1985,Clance1978}. It is thought to affect high-achieving individuals across various domains, including higher education and high-skill professions \cite{Chakraverty2022,Trinkenreich2022}. Although there have been studies examining how women and minority groups cope with imposter syndrome \cite{maji_they_2021}, and how other variables are correlated such as self-esteem \cite{naser_impostor_2022}, and self-efficacy \cite{pakozdy_imposter_2023}, there is limited research on the connection between impostor syndrome and undergraduate computer science students.

As undergraduate students progress through their computer science degree, they not only acquire new skills but also refine existing ones, ultimately becoming proficient in designing and implementing software systems. By the time they reach the end of their academic journey, these students possess the qualifications necessary to begin their careers in the tech industry. It is natural to assume that at this stage, these students are confident in their skills and feel well-prepared for professional success. However, previous research indicates that students enrolled in STEM programs, including those related to computer science, often experience feelings of inadequacy, potentially impacting their confidence in their skills \cite{Rosenstein2020}. As such, this study aims to examine the prevalence and relationship of impostor syndrome with code comprehension performance in fourth-year computer science students through an augmented cognition approach.

\subsection{Goal \& Research Questions}
Code comprehension is an essential activity in software development and maintenance. Source code conveys the system's behavior, through identifier names \cite{Peruma2018}, which developers rely on to understand the code they are working on to fix defects or incorporate feature changes \cite{Mayrhauser1995Computer}. As computer science students prepare to enter the workforce, feelings of self-doubt and insecurity associated with impostor syndrome can undermine their ability to comprehend codebases, thereby negatively impacting their confidence and job performance. Therefore, this study aims to understand how impostor syndrome impacts code comprehension. We envision our study laying the foundation for future research in this area while also providing initial insights to improve computer science education and support students as they transition to professional roles. To this extent, we aim to answer the following research questions (RQs):

\vspace{1mm}
\noindent\textbf{RQ 1: To what extent are final-year undergraduate computer science students confident in their program comprehension skills?} This research question intends to examine the level of impostor syndrome exhibited by computer science students who are about to graduate. The research question assesses the demographic factors closely associated with impostor syndrome.

\vspace{1mm}
\noindent\textbf{RQ 2: How does impostor syndrome affect cognitive processes involved in comprehending code?} This research question aims to better understand cognitive processes involved in comprehending code, and how impostor syndrome can affect these processes. This is achieved by tracking eye movements, monitoring heart rates, and taking other measurements, such as the confidence of participants, time taken to complete a code question, and how often participants answered the coding question correctly.

\subsection{Contribution}
The main contributions from this work are as follows:
\begin{itemize}
\item We provide preliminary yet promising findings on the extent to which impostor syndrome affects the code comprehension cognition of experienced undergraduate students.
\item This study establishes groundwork for further research exploring interventions against impostor syndrome.
\item We make our dataset publicly accessible for replication/extension purposes.
\end{itemize}

 \subsection{Paper Structure}
The rest of this paper is organized as follows: Section \ref{Section:related} presents a review of related work on imposter syndrome in software engineering. Section \ref{Section:design} describes the methodology used for this study. Section \ref{Section:results} provides answers to both research questions and reports on the results of the proposed experiments. Section \ref{Section:threats} discusses the potential threats to the validity of the study. Finally, Section \ref{Section:conclusion} concludes by summarizing the findings and suggesting potential directions for future work.

\section{Related Work}
\label{Section:related}

In this section, we report on related work in this area. While there exist studies that examine imposter syndrome in students and industry professionals, we limit our focus to the literature that addresses imposter syndrome among students who are either taking a computing course or studying a computing subject/topic, as well as software engineering professionals. 

In a survey with 200 undergraduate computer science students, Rosenstein et al. \cite{Rosenstein2020} show that 57\% of the surveyed students exhibited frequent feelings of the Impostor Phenomenon. The authors also highlight that women experienced these feelings more frequently (71\%) compared to men (52\%). Heels and Devlin \cite{Heels2019} examine the roles chosen by female students in a large software engineering team project and report that despite strong academic backgrounds, female students tend to opt for less technical roles than male students. The authors recommend exploring how unconscious bias or imposter syndrome affects female students. Interviews with 16 Digital Technologies teachers from primary and secondary schools by Varoy et al. \cite{Varoy2023} show that female students experience imposter syndrome more than male students and feel they are not making as much progress as male students, even when they are doing better work. The teachers note that male students tend to loudly proclaim their achievements, while female students work more quietly and cooperatively. This led the female students to underestimate their own abilities and progress. In a study on coding bootcamps, Thayer and Ko \cite{Thayer2017} found that imposter syndrome can act as an informal boundary to entering the software industry and persist even after gaining experience and employment. Additionally, participants in coding bootcamps can also experience imposter syndrome, similar to those in the software industry.

A survey by Ginter \cite{ginter2023imposter} of 104 software engineers shows that individuals exhibiting feelings of imposter syndrome usually had insecure attachment styles, particularly anxiety and avoidance. The study also found that anxious attachment styles in individuals with imposter syndrome were linked to higher levels of depression and anxiety. In a study that surveyed 94 women employed at a global technology company, Trinkenreich et al. \cite{Trinkenreich2022ICSE} note that imposter syndrome was a challenge women faced in software development teams and could lead to them leaving the organization. In an online survey of 134 Finnish women, Hyrynsalmi and Hyrynsalmi \cite{Hyrynsalmi2019} reports that women are motivated to transition to the software industry but encounter challenges such as a lack of career counseling, uncertainty about the required education, and issues related to self-esteem and imposter syndrome.

\section{Study Design}
\label{Section:design}
In this section, we provide details about the design of this study. At a high level, the study consisted of participants reviewing a set of code snippets and answering questions through online questionnaires. While reading code snippets, participants' eye movements and heart rates were monitored. We will elaborate on these activities below. The experiment took place in a private room without any windows, where only the participant and one investigator (i.e., an author) were present. A 24-inch monitor was used to display code snippets and questionnaires. The investigator used a secondary monitor, mouse, and keyboard to manage the experiment. After the calibration stage and before starting the experiment, each participant went through a trial run. This allowed to check if the eye tracker and biometric devices were functioning adequately but also to make sure participants understood and got familiar with the task. A couple trial code snippets were then displayed during this process and have not been included in the results. This study was approved by the Institutional Review Board of our institute and all participants provided consent prior to participating. The high-resolution images of the code snippets, survey questionnaire, and participant data are available at: \cite{ProjectWebsite}.

\subsection{Survey Design}
As part of our study, we used three types of online questionnaires - a demographic pre-questionnaire, a code snippet questionnaire, and an impostor syndrome post-questionnaire. To construct and host these questionnaires, we leveraged Qualtrics. All participants answered survey questions using the same workstation that was connected to the eye tracker and biometrics device. Below, we have provided a brief description of each questionnaire.

Prior to code comprehension activities and questions, participants completed a pre-questionnaire capturing their demographics. All questions were required and are shown in Table \ref{Table:SurveyQuestionsPre}.

\begin{table}
\centering
\vspace{-15pt}
\caption{Pre-Questionnaire capturing demographic details.}
\label{Table:SurveyQuestionsPre}
\scriptsize
\begin{tabular}{lp{0.8\linewidth}p{0.2\linewidth}}
\toprule
\textbf{No.} &
  \textbf{Question} &
  \textbf{Type} \\ \midrule\midrule
1 &
  What best describes your gender? &
  Single-Choice \\ \midrule
2 &
  What best describes your ethnicity? &
  Multi-Choice \\ \midrule
3 &
  What is your current academic year? &
  Single-Choice \\ \midrule
4 &
  What is your primary field of study/major? &
  Single-Choice \\ \midrule
5 &
  Are you a first generation college student? &
  Single-Choice \\ \midrule
6 &
 Do you have an immediate family member who has worked or is working in the software industry? (this includes internships) &
  Single-Choice\\ \midrule
7 &
  How many years of experience in the software industry do you have? (this includes internships) &
  Single-Choice \\ \midrule
8 &
  How many years of Java programming experience do you have? &
  Single-Choice \\ \midrule
9 &
  What is your preferred programming language? &
  Single-Choice  \\ \midrule
10 &
  Do you have a software engineering-related job lined up post-graduation? &
  Single-Choice \\ \midrule
11 &
  How would you rate your overall experience with your programming education?
 &
  Single-Choice   \\ \bottomrule
\end{tabular}
\vspace{-15pt}
\end{table}

During the coding analysis task, participants were presented with five individual code snippets, each accompanied by one single-choice question. The question required participants to either identify the correct output of the code or identify the number of the line where the code was causing a logical, runtime, or compile-time error. Participants only had five minutes to read each code snippet and answer the associated question. To reduce the possibility of participants guessing the answer, each question included an ``I don't know'' option. Section \ref{Section:results} provides screenshots of each code snippet, along with their associated question and multiple options for answer. Right after participants provided an answer to the question related to a specific code snippet, they were asked to provide feedback on the confidence-level for the answer they selected.
After completing code snippets' comprehension activities, participants were asked to fill out a post-questionnaire to assess the extent of their imposter syndrome. For this study, Clance IP Scale questions \cite{Clance1985} were customized to focus only on source code analysis and troubleshooting \footnote{From The Impostor Phenomenon:  When Success Makes You Feel Like A Fake (pp. 20-22), by P.R. Clance, 1985, Toronto:  Bantam Books.  Copyright 1985 by Pauline Rose Clance.  Reprinted by permission.  Do not reproduce without permission from Pauline Rose Clance, drpaulinerose@comcast.net.}. Those questions can be found in Table \ref{Table:SurveyQuestionsPost}. All participants were required to answer all questions in this questionnaire.

\begin{table}
\vspace{-15pt}
\centering
\caption{Post-Questionnaire containing the customized Clance IP Scale questions that focus on source code analysis and troubleshooting.}
\label{Table:SurveyQuestionsPost}
\scriptsize
\begin{tabular}{lp{0.8\linewidth}p{0.2\linewidth}}
\toprule
\textbf{No.} &
  \textbf{Question} &
  \textbf{Type} \\ \midrule\midrule
1 &
  I have often succeeded in programming tasks, even though I was afraid that I would not do well before I started working on it. &
  Single-Choice \\ \midrule
2 &
  I can give the impression that I'm more competent in my programming skills than I really am. &
  Single-Choice \\ \midrule
3 &
  I tend to avoid programming tasks and have a sense of dread when others assess my programming work. &
  Single-Choice \\ \midrule
4 &
  When people praise me for my code analysis and troubleshooting abilities, I'm afraid I won't be able to live up to their expectations of me in the future. &
  Single-Choice \\ \midrule
5 &
  I sometimes think my success in code analysis and troubleshooting is due to external factors (i.e., environmental or people) rather than due to my skills. &
  Single-Choice \\ \midrule
6 &
 I'm afraid people important to me may find out that I'm not as capable at programming as they think I am. &
  Single-Choice\\ \midrule
7 &
  I tend to remember the incidents in which I have not done my best in programming more than those times I have done my best. &
  Single-Choice \\ \midrule
8 &
  I rarely do a programming task as well as I'd like to do it. &
  Single-Choice \\ \midrule
9 &
  Sometimes I feel or believe that my success in analyzing and troubleshooting code has been the result of some kind of accident. &
  Single-Choice  \\ \midrule
10 &
  It's hard for me to accept compliments or praise about my programming skills or accomplishments &
  Single-Choice \\ \midrule
11 &
  At times, I feel my success in code analysis and troubleshooting has been due to some kind of luck. &
  Single-Choice \\ \midrule
12 &
  I'm disappointed at times in my code analysis and troubleshooting accomplishments and think I should have accomplished much more. &
  Single-Choice  \\ \midrule
13 &
  Sometimes I'm afraid others will discover how much code analysis and troubleshooting knowledge or ability I really lack. &
  Single-Choice \\ \midrule
14 &
  I'm often afraid that I may fail at a new code analysis and troubleshooting assignment or undertaking, even though I generally do well at what I attempt. &
  Single-Choice \\ \midrule
15 &
  When I've succeeded at programming and received recognition for my accomplishments, I have doubts that I can keep repeating that success. &
  Single-Choice \\ \midrule
16 &
  If I receive a great deal of praise and recognition for my code analysis and troubleshooting accomplishments, I tend to discount the importance of what I've done. &
  Single-Choice \\ \midrule
17 &
  I often compare my code analysis and troubleshooting abilities to those around me and think they may be more intelligent than I am. &
  Single-Choice \\ \midrule
18 &
  I often worry about not succeeding with a programming task, even though others around me have considerable confidence that I will do well. &
  Single-Choice \\ \midrule
19 &
  If I'm going to gain recognition for my code analysis and troubleshooting skills, I hesitate to tell others until it is an accomplished fact. &
  Single-Choice  \\ \midrule
20 &
  I feel bad and discouraged if I'm not "the best" or at least "very special" in code analysis and troubleshooting situations that involve achievement. &
  Single-Choice  \\ \bottomrule
\end{tabular}
\vspace{-15pt}
\end{table}

\subsection{Code Snippets}
In this study, each participant was required to review five code snippets and answer their associated question. Code snippets were in the Java programming language as it is widely used in multiple courses at the University, and all students are familiar with this programming language. Code snippets covered concepts such as data structures, recursion, sorting, and string analysis, which are taught in undergraduate-level courses and are usually asked during job interviews in industry. Table \ref{Table:codeDescription} provides a summary of code snippets. The code snippets and their associated question are shown in Section \ref{Section:results}.  

\begin{table}
\vspace{-15pt}
\centering
\caption{Description of the five code snippets utilized in this study.}
\label{Table:codeDescription}
\resizebox{\columnwidth}{!}{%
\scriptsize
\begin{tabular}{lp{0.7\linewidth}p{0.3\linewidth}}
\toprule
\multicolumn{1}{c}{\textbf{Id}} &
  \multicolumn{1}{c}{\textbf{Description}} &
  \multicolumn{1}{c}{\textbf{Task}} \\ \midrule
1 &
  This code snippet involves array size manipulation within a loop that will cause a runtime issue. Unlike the error in the other code snippets, participants were not explicitly informed that an error exists in the code. &
  Determine the output \\ \midrule
2 &
  This is an implementation of the bubble sort algorithm. However, there is an error in the condition used in the loop, which will lead to the list being sorted incorrectly. The method's name is called `bubbleSort', so the participant is aware of the intended behavior of the code. &
  Identify the number of the line where the error occurs \\ \midrule
3 &
  This is a recursive function that prints a sequence of digits. There are no errors in the code. &
  Determine the output \\ \midrule
4 &
  This code snippet accepts a provided string and prints the length of the last word in the provided string. The names of the identifiers in the code do not indicate the code's behavior. There are no errors in this code. &
  Determine the output \\ \midrule
5 &
  This code checks if a given string is a palindrome. The implementation contains an error in the calculation that results in a runtime error. &
  Identify the number of the line where the error occurs \\ \bottomrule
\end{tabular}
}\vspace{-15pt}
\end{table}

\subsection{Eye Tracker \& Biometric Device}
Eye movements and physiological samples such as heart rate were recorded respectively using the Gazepoint GP3 HD Eye Tracker and the Biometrics device. These devices are research-grade and are commonly used for academic as well as medical research \cite{Gazepoint,Gazepoint2}. The eye tracker has a sampling rate of 150 Hz to reduce the chance of loss of tracking and a 0.5 – 1.0 degree of visual angle accuracy. During the experiment, participants sat on a chair facing a computer monitor where code snippets and surveys were presented to them. Prior to the experiment, the camera was calibrated using the Gazepoint Control software. Participants were asked to make saccades to nine targets presented on the screen for calibration purposes. Those nine targets were presented at the top, at the middle and at the bottom of the screen going from the left through the middle to the right of the screen.
The heart rate was measured using a finger sensor module. The Gazepoint Analysis software was utilized to manage the experiment, including starting/stopping the eye tracker and the creation of areas of interest (i.e., AOIs).

\subsection{Data Preprocessing}
Once data collection was completed, data for analysis was prepared using Python. The temporal gaze point data (a single gaze point represents where a user looked and when in milliseconds) from Gazepoint was first cleaned by removing durations of time where: (1) the code was not on the screen due to latency from Qualtrics when displaying the survey question, (2) the time between code snippets, and (3) the participant was looking away from the screen and was not focused on the code snippet. The times for (1) and (3) were determined manually by two investigators watching the video recordings with precision up to three decimal places. The cutoff time for (2) was automated by trimming the data up to when the participant submitted their answer for a code snippet question.

The following aggregate measures were calculated for each participant: average heart rate, total time spent on a snippet, and most looked at AOI (specifically, only AOIs containing code were analyzed for this study; answer choices below each code snippet were not considered). Duration spent in each AOI was determined by summing the differences in time between the first instance when a gaze point was identified to be within an AOI and the first instance when the participant moved to a different AOI. All durations for the AOIs were then normalized by the number of words, excluding non-alphanumeric characters (as described in Sharafi et al.\cite{sharafi2020practical}). 

Other calculations for the data included determining levels of imposter syndrome (detailed in RQ1 of Section \ref{Section:results}) and determining the correctness of answers (a binary true/false if the answer was correct) for each code snippet which were determined by two investigators. Additionally, recoding of ordinal variables to be numerical were coded starting from the number zero and non-codeable responses were designated as NaN.

\subsection{Participants}
In this study, fifteen final-year undergraduate students enrolled in the Computer Science program at the Information and Computer Sciences Department of the University of Hawaiʻi at Mānoa were recruited. These participants were not given any monetary compensation or extra credit for participating in the study; participation was voluntary. 

Out of the fifteen participants who took the survey, nine of them identified as male, five as female, and one as non-binary/third gender. Moreover among the participant pool, ten identified themselves as Asian only, one as White/Caucasian, one as Black/African American, and the remaining three participants identified with two or more ethnicities. 

In terms of Java programming experience, eight participants indicated they had one to two years of experience, four participants mentioned they had three to five years of experience, and three participants reported they had less than one year of experience. Additionally, six participants had between one to two years of industry experience, five participants had less than a year of industry experience, two participants had three to five years of industry experience, and the remaining two participants had no prior industry experience. Table \ref{Table:demographics} provides a breakdown of participants' demographics.

\begin{table}[]
\caption{Demographic details of the participants in the study.}
\label{Table:demographics}
\resizebox{\columnwidth}{!}{%
\begin{tabular}{lllll}
\hline
\textbf{\begin{tabular}[c]{@{}l@{}}Participant\\ ID\end{tabular}} &
  \textbf{Gender} &
  \textbf{Ethnicity} &
  \textbf{\begin{tabular}[c]{@{}l@{}}Years of Exp. \\ in Software \\ Industry\end{tabular}} &
    \textbf{\begin{tabular}[c]{@{}l@{}}Years of Exp. \\ in Java\end{tabular}} \\ \midrule\midrule
1  & Male   & White or Caucasian        & 1-2 years            & 1-2 years            \\ \hline
2  & Female & Asian                     & 1-2 years            & 1-2 years            \\ \hline
3  & Female & Asian, White or Caucasian  & 1-2 years            & 3-5 years            \\ \hline
4  & Male   & Asian                     & None                 & 3-5 years            \\ \hline
5  & Male   & Asian                     & \textgreater{}1 year & 3-5 years            \\ \hline
6  & Female & Asian                     & 1-2 years            & 1-2 years            \\ \hline
7  & Female & Asian                     & 3-5 years            & 3-5 years            \\ \hline
8  & \begin{tabular}[c]{@{}l@{}}Non-binary / \\ third gender\end{tabular} &
  Asian &
  \textgreater{}1 year &
  \textgreater{}1 year \\ \hline
9  & Male   & Black or African American & None                 & 1-2 years            \\ \hline
10 & Male   & Asian                     & 1-2 years            & \textgreater{}1 year \\ \hline
11 & Female & Asian, Native Hawaiian     & 1-2 years            & 1-2 years            \\ \hline
12 & Male   & Asian                     & \textgreater{}1 year & 1-2 years            \\ \hline
13 & Male   & Asian                     & 3-5 years            & 1-2 years            \\ \hline
14 & Male   & Asian                     & \textgreater{}1 year & \textgreater{}1 year \\ \hline
15 &
  Male & \begin{tabular}[c]{@{}l@{}} American Indian or Alaska Native, \\ Asian, White or Caucasian\end{tabular}  &
  \textgreater{}1 year &
  1-2 years \\ \hline
\end{tabular}
}
\end{table}

\subsection{Pilot Study}
We followed the best practice of conducting a pilot study to identify any flaws in our methodology. For this purpose, we recruited four students to participate in a pilot study, who were not included in the actual study. All data collected during the pilot study were discarded after its completion. During the pilot run, we identified certain survey questions that needed to be reworded to improve clarity as well as code snippets that required consistent formatting. We also made improvements to the positioning of the eye tracker and the participant's chair. Furthermore, the pilot study helped us construct a script/instructions to follow when conducting the actual study.

\section{Results}
\label{Section:results}
In this section, we report the study's results by answering our RQs.

\subsection{\noindent\textbf{RQ 1: To what extent are final-year undergraduate computer science students confident in their program comprehension skills?}}

\begin{wraptable}{r}{0.3\textwidth}
\vspace{-10pt}
\vspace{-\baselineskip}
    \caption{Range of values associated with each IPC level}\label{wrap-tab:1}
    \begin{tabular}{ll}
    \hline
    \textbf{IPC Level} & \textbf{Scores} \\ \hline
    Few                & \textless  40   \\ \hline
    Moderate           & 41 - 60         \\ \hline
    Frequent           & 61 - 80         \\ \hline
    Intense            & \textgreater 80 \\ \hline
    \end{tabular}
\vspace{-\baselineskip}
\vspace{-10pt}
\end{wraptable} 
Based on the Clance Imposter Phenomenon (IP) Scale \cite{Clance1985}, participants were sorted into four Imposter Phenomenon Characteristics (IPC) categories: Few IPC for those who demonstrated few imposter syndrome characteristics, Moderate IPC for those with moderate characteristics, Frequent IPC, and Intense IPC, with each corresponding to the severity level of IP. These levels correspond to how frequently and seriously IP interferes with a person's life. 

Once participants filled out the post-questionnaire based on a version of the Clance IP Scale adapted to the task, their score was calculated as follows: Each question had a range of multiple-choice answers, which were "not at all true," "rarely," "sometimes," "often," and "very true." Each answer mapped to a numerical value from 1 to 5, respectively. These values are then tallied to create an aggregate score for each participant and used to classify each participant into one of the four IPC groups. The range of values associated with each IPC category can be found in Table \ref{wrap-tab:1}.

Of the fifteen participants, three had Few IPC, five had Moderate IPC, five had Frequent IPC, and two had Intense IPC. Of the three participants with Few IPCs, all three identified themselves as male. In the Moderate and Frequent IPC groups, each comprising five individuals, two participants identified themselves as female and three as male. Amongst the two participants with Intense IPC, one identified as female and the other one as non-binary/third gender. 

A t-test to compare the IPC scores between males and females (the participant who responded "non-binary" was omitted) found that the nine males (M = 50.56, SD = 12.32) compared to the five females (M = 68.80, SD = 12.28) scored significantly lower on the modified Clance IP scale, t(12) = -2.658, p = .021. Therefore, females were associated with higher levels of the imposter phenomenon, particularly in relation to source
code analysis and troubleshooting.

One-way ANOVA tests were additionally conducted to compare how IPC scores were related to software industry experience and Java experience. Findings show that there is no significant difference in IPC scores so there is no relationship between these variables (p > .05 for both).

\begin{tcolorbox}[top=0.5pt,bottom=0.5pt,left=1pt,right=1pt]
\textbf{Summary for RQ1.}
We found that female students had higher characteristics of imposter syndrome. The number of years of experience in the software industry or the tested programming language did not seem to correlate to the level of characteristics of imposter syndrome.
\end{tcolorbox}
\vspace{-15pt}

\subsection{\noindent\textbf{RQ 2: How does imposter syndrome affect the cognitive processes involved in comprehending code?}}

\begin{figure}[H]
\vspace{-10pt}
    \centering
    \setlength{\abovecaptionskip}{0pt} 
    \setlength{\belowcaptionskip}{0pt} 
    
    \includegraphics[width=\textwidth]{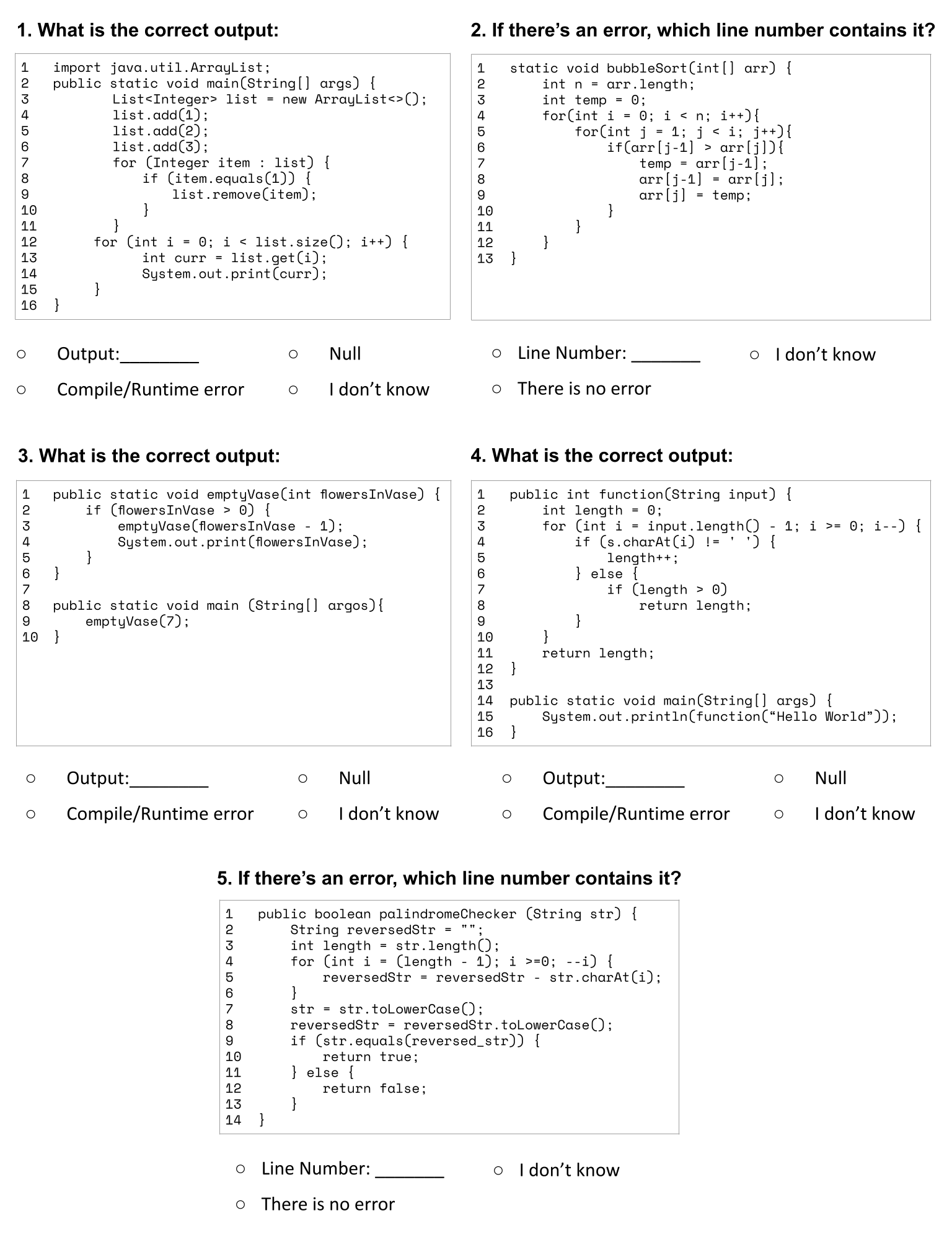}
    
    \caption{The five code snippets that were part of our study along with their associated question and answer choices.}
    \label{fig:all_snippets}
\end{figure}

To answer this research question, we examine the eye tracking and biometrics results of participants comprehending each code snippet in the study. Below, for each code snippet, we discuss the average heart rate, Areas of Interest (AOI) metrics, time spent by participants on the snippet, the percentage of correct answers, and the extent to which participants are confident with their answers. Additionally, Figure \ref{fig:all_snippets} displays all five code snippets that were included in our study, along with their corresponding question.

AOIs are regions of a stimulus from which quantitative metrics can be derived \cite{sharafi2020practical}. Each AOI contains its own set of metrics, e.g., the average time participants took to look at a specific AOI. Each line of code, excluding those solely comprising of curly braces, was designated an AOI. Each stimulus had an average of 13.4 AOIs. As mentioned in Section \ref{Section:design}, each AOI was normalized by the number of words it contained on its line in order to allow fair comparison between AOIs (words were characterized as sequences of alphanumeric characters that are separated by spaces after excluding non-alphanumeric characters). 

The AOIs were then grouped into one of ten different categories, depending on the functionality of the line of code. These categories are as follows: else statement, for loop, if statement, import statement, method call, method declaration, recursive method call, return statement, variable assignment, and variable declaration. An example of a method declaration is ``public static void main(String[] args) \{'', while an example of a method call is ``list.add(1).''

\noindent{\textbf{Code Snippet 1 -  Array Size Manipulation}}

\snippetResults{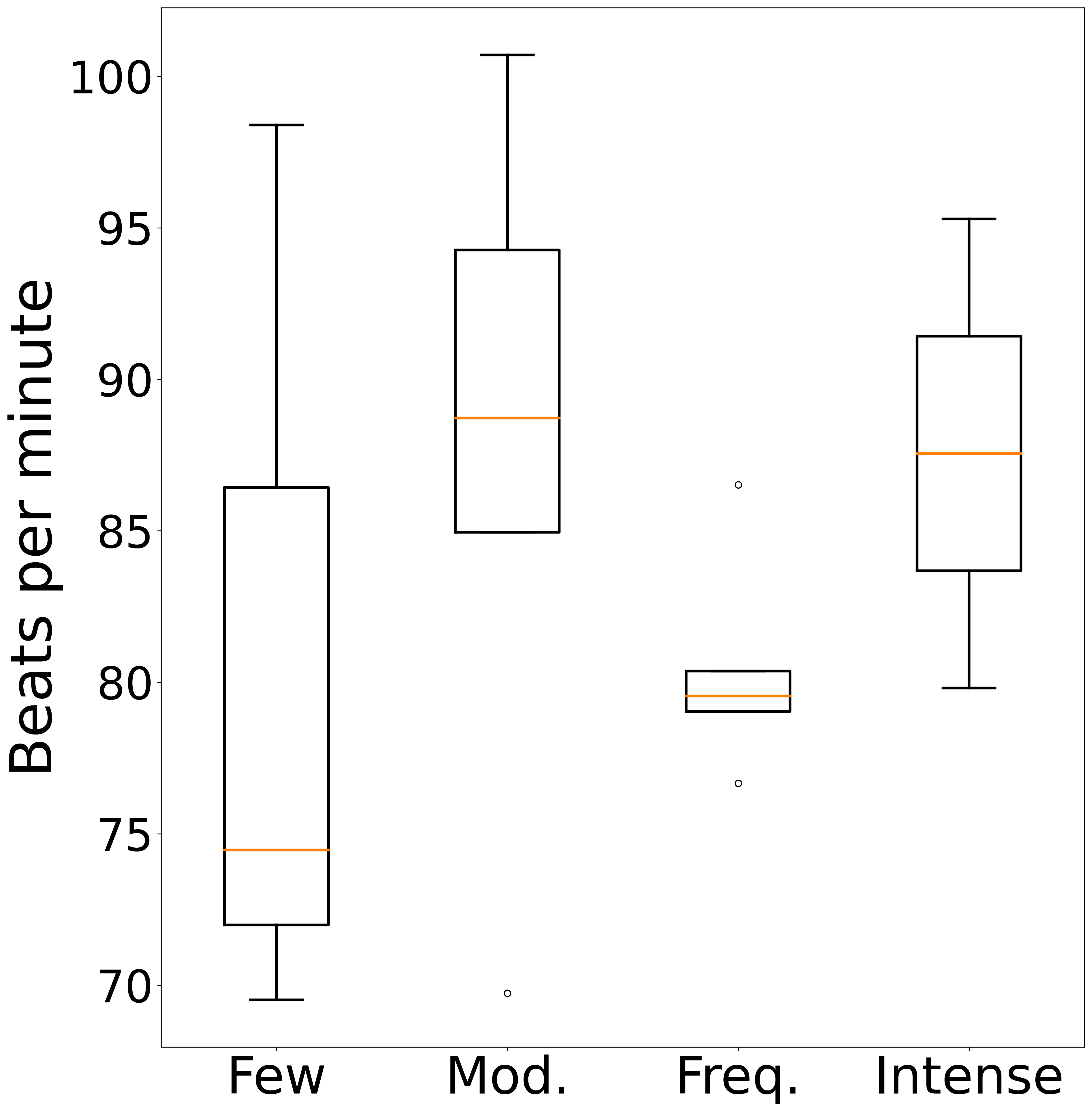}
It was found that for Code Snippet 1, only one participant from the Moderate IPC group answered the problem correctly (see Figure \ref{fig:correctlyanswered_1}). The lowest median for the average heart rate occurred in the Few IPC group, whereas the highest median average heart rate occurred in the Moderate IPC group (see Figure \ref{fig:avgheartrate_1}). The amount of time participants took before submitting an answer was the lowest for those in the Few IPC group and the highest in the Intense IPC group (see Figure \ref{fig:timespent_1}).

The lines of code that participants spent the longest duration on were import statements for Few IPC, Moderate IPC, and Frequent IPC (for each row in Table \ref{tab:Snippet1_aois}, the percentage was highest for these three IPC groups). Those in the Intense IPC spent the longest duration on if statements (23.87\%). For all but those in Frequent IPC, the least amount of time was spent on method declarations.
\begin{table}
\vspace{-15pt}
\caption{Duration spent on a Code Category in Snippet 1}
\label{tab:Snippet1_aois}
\renewcommand{\arraystretch}{1.5}
\resizebox{\columnwidth}{!}{%
\scriptsize
\begin{tabular}{r|*{8}{>{\centering\arraybackslash}m{1.7cm}|}}
\cline{2-8}
\multicolumn{1}{l|}{} &
  \textbf{For Loop} &
  \textbf{If Stmt} &
  \textbf{Import Stmt} &
  \textbf{Method Call} &
  \textbf{Method Decl} &
  \textbf{Var Decl} &
  \textbf{Multiple} \\ \hline
\multicolumn{1}{|r|}{\textbf{Few IPC}} &
  \vspace{3pt}16.18\% (0.46)\vspace{3pt} &
  \vspace{3pt}18.28\% (0.52)\vspace{3pt} &
  \cellcolor[HTML]{E06666}\vspace{3pt}21.93\% (0.62)\vspace{3pt} &
  \vspace{3pt}21.29\% (0.6)\vspace{3pt} &
  \cellcolor[HTML]{B7D7A8}\vspace{3pt}10.71\% (0.3)\vspace{3pt} &
  \vspace{3pt}11.14\% (0.32)\vspace{3pt} &
  \vspace{3pt}0.47\% (0.01)\vspace{3pt} \\ \hline
\multicolumn{1}{|r|}{\textbf{Moderate IPC}} &
  \vspace{3pt}17.17\% (0.8)\vspace{3pt} &
  \vspace{3pt}12.7\% (0.59)\vspace{3pt} &
  \cellcolor[HTML]{E06666}\vspace{3pt}25.17\% (1.17)\vspace{3pt} &
  \vspace{3pt}18.25\% (0.85)\vspace{3pt} &
  \cellcolor[HTML]{B7D7A8}\vspace{3pt}11.62\% (0.54)\vspace{3pt} &
  \vspace{3pt}14.51\% (0.67)\vspace{3pt} &
  \vspace{3pt}0.58\% (0.03)\vspace{3pt} \\ \hline
\multicolumn{1}{|r|}{\textbf{Frequent IPC}} &
  \vspace{3pt}17.76\% (0.91)\vspace{3pt} &
  \cellcolor[HTML]{B7D7A8}\vspace{3pt}10.93\% (0.56)\vspace{3pt} &
  \cellcolor[HTML]{E06666}\vspace{3pt}30.31\% (1.56)\vspace{3pt} &
  \vspace{3pt}16.01\% (0.82)\vspace{3pt} &
  \vspace{3pt}12.66\% (0.65)\vspace{3pt} &
  \vspace{3pt}12.07\% (0.62)\vspace{3pt} &
  \vspace{3pt}0.24\% (0.01)\vspace{3pt} \\ \hline
\multicolumn{1}{|r|}{\textbf{Intense IPC}} &
  \vspace{3pt}20.63\% (1.18)\vspace{3pt} &
  \cellcolor[HTML]{E06666}\vspace{3pt}23.87\% (1.36)\vspace{3pt} &
  \vspace{3pt}14.71\% (0.84)\vspace{3pt} &
  \vspace{3pt}17.65\% (1.01)\vspace{3pt} &
  \cellcolor[HTML]{B7D7A8}\vspace{3pt}9.67\% (0.55)\vspace{3pt} &
  \vspace{3pt}11.26\% (0.64)\vspace{3pt} &
  \vspace{3pt}2.21\% (0.13)\vspace{3pt} \\ \hline
\end{tabular}
}
\medskip
\scriptsize\\
Row percentages for each IPC are shown in each cell with red highlights denoting largest percentage and green highlights denoting smallest (ignoring the Multiple category). The duration normalized by number of words are in parenthesis. 
\vspace{-15pt}
\end{table}

\pagebreak\noindent{\textbf{Code Snippet 2 - Bubble Sort}}

\snippetResults{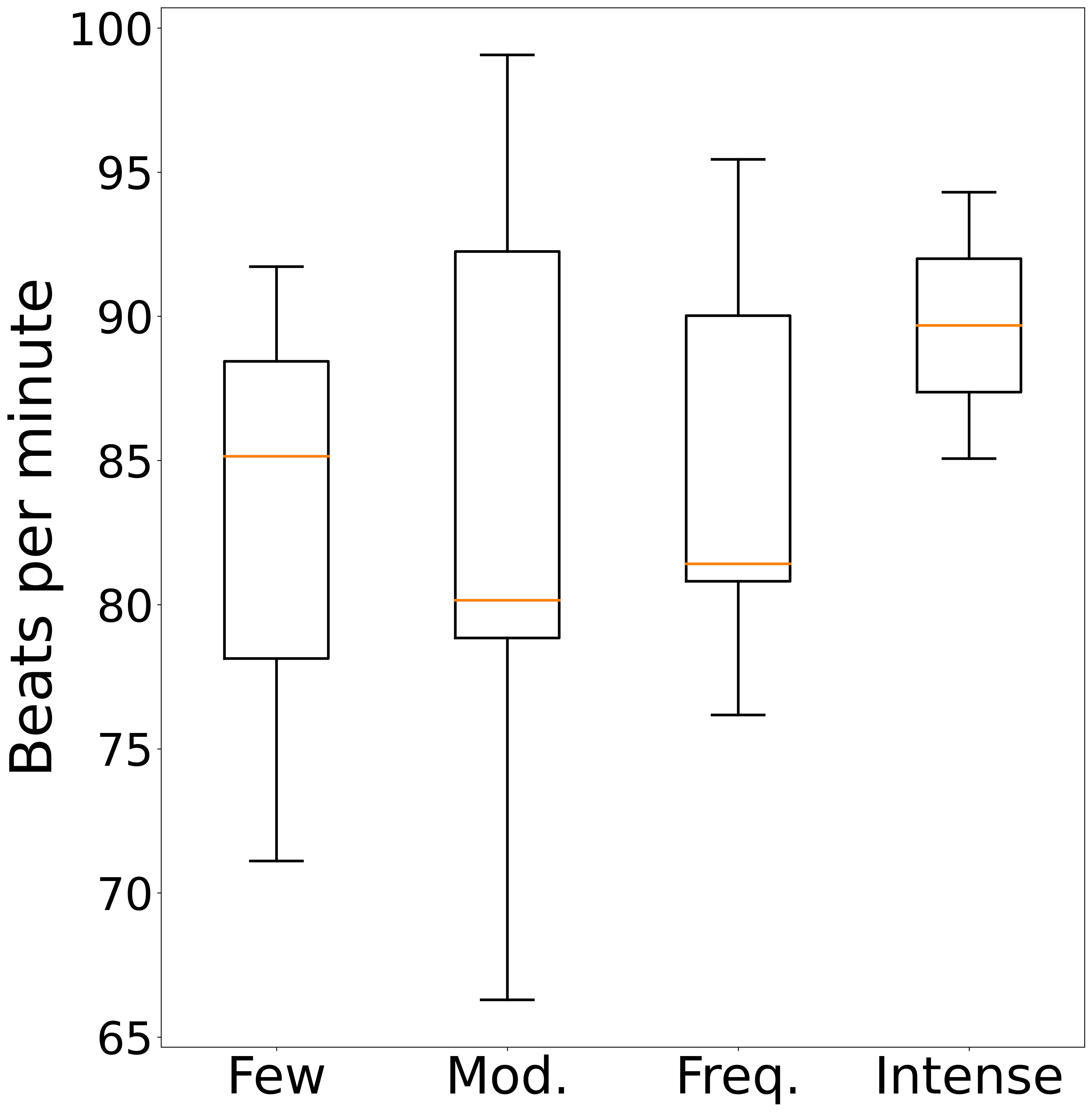}
As shown in Figure \ref{fig:biometrics_other_per_snippet_2}, only two participants from the Frequent IPC group answered the problem correctly. The average heart rate was the lowest in the Moderate IPC group and Highest in the intense IPC group. Participants in the Few IPC group spent the least amount of time and those in the Intense IPC group spent the most.

Referring to Table \ref{tab:Snippet2_aois}, those from the Few IPC, Moderate IPC, and Frequent IPC spent the least amount of time on the if statements, while those in the Intense IPC group spent the least amount of time on variable assignments. Besides the Moderate IPC group, every group spent the most time on method declarations.
\begin{table}
\vspace{-15pt}
\caption{Duration spent on a Code Category in Snippet 2}
\label{tab:Snippet2_aois}
\renewcommand{\arraystretch}{1.1} 
\resizebox{\columnwidth}{!}{
\scriptsize
\begin{tabular}{r|*{6}{>{\centering\arraybackslash}m{1.5cm}|}} 
\cline{2-6}
\multicolumn{1}{l|}{}                      & \textbf{For Loop} & \textbf{If Stmt} & \textbf{Method Decl}                   & \textbf{Var Asgmt} & \textbf{Var Decl} \\ \hline
\multicolumn{1}{|r|}{\textbf{Few IPC}} &
   14.09\% (0.66)  &
  \cellcolor[HTML]{B7D7A8} 8.33\% (0.39)  &
  \cellcolor[HTML]{E06666} 39.24\% (1.84)  &
   11.82\% (0.55)  &
   26.52\% (1.24)  \\ \hline
\multicolumn{1}{|r|}{\textbf{Moderate IPC}} &
   16.45\% (1.01)  &
  \cellcolor[HTML]{B7D7A8} 13.05\% (0.8)  &
   24.17\% (1.48)  &
  \cellcolor[HTML]{E06666} 25.87\% (1.59)  &
   20.46\% (1.26)  \\ \hline
\multicolumn{1}{|r|}{\textbf{Frequent IPC}} &
   19.54\% (1.69)  &
  \cellcolor[HTML]{B7D7A8} 10.92\% (0.95)  &
  \cellcolor[HTML]{E06666} 36.96\% (3.2)  &
   12.79\% (1.1)  &
   19.89\% (1.72)  \\ \hline
\multicolumn{1}{|r|}{\textbf{Intense IPC}} &  17.95\% (1.01)     &  13.02\% (0.73)    & \cellcolor[HTML]{E06666} 36.51\% (2.06)  &  \cellcolor[HTML]{B7D7A8} 12.7\% (0.72)       &  19.74\% (1.11)    \\ \hline
\end{tabular}
}
\vspace{-15pt}
\end{table}

\pagebreak\noindent{\textbf{Code Snippet 3 - Recursion}}

\snippetResults{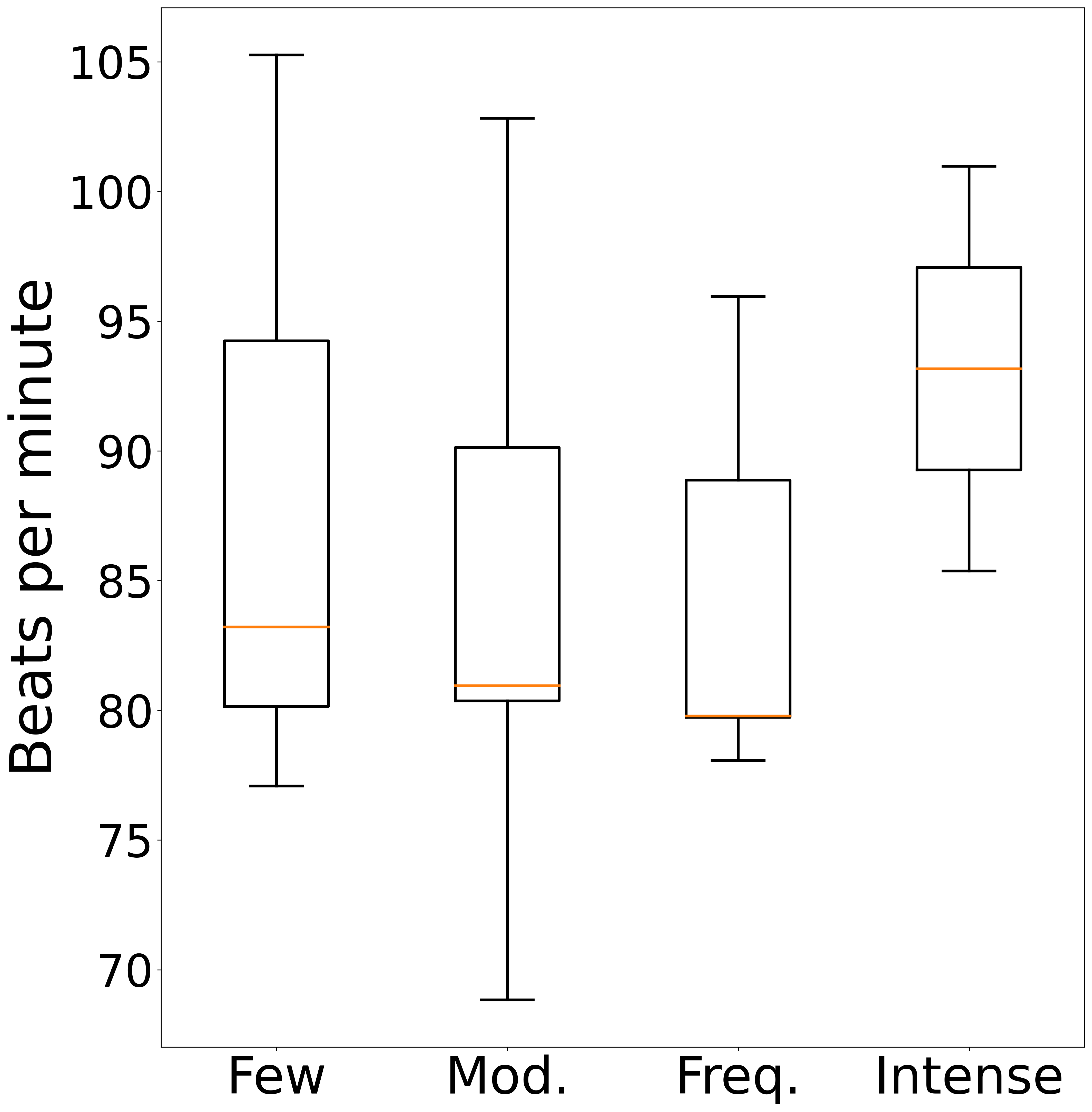}
For Code Snippet 3, all of the participants in the Few IPC group answered the problem correctly, whereas no participants from the Intense IPC group answered correctly (Figure \ref{fig:biometrics_other_per_snippet_3}). One participant from the Moderate and Frequent IPC groups answered correctly. The lowest median average heart rate was from the Frequent IPC group, and the highest median was from the Intense IPC group. Participants from the Intense IPC group spent the longest amount of time on the snippet compared to those in the Few IPC group, who spent the shortest.

This snippet had less consistent results across the IPC groupings as can be seen in Table \ref{tab:Snippet3_aois}. However, it can seen that those in the Few IPC group spent the most time on if-statements and the least amount of time on recursive calls, whereas those in the Intense IPC group spent most of their time on method declarations and the least amount of time on if-statements.
\begin{table}
\vspace{-15pt}
\caption{Duration spent on a Code Category in Snippet 3}
\label{tab:Snippet3_aois}
\renewcommand{\arraystretch}{1.2} 
\resizebox{\columnwidth}{!}{ 
\scriptsize
\begin{tabular}{r|*{5}{>{\centering\arraybackslash}m{1.6cm}|}} 
\cline{2-5}
\multicolumn{1}{l|}{} & \textbf{If Stmt} & \textbf{Method Call} & \textbf{Method Decl} & \textbf{Recursive Call} \\ \hline
\multicolumn{1}{|r|}{\textbf{Few IPC}} &
  \cellcolor[HTML]{E06666}31.25\% (1.08) &
  22.73\% (0.79) &
  30.88\% (1.07) &
  \cellcolor[HTML]{B7D7A8}15.14\% (0.52) \\ \hline
\multicolumn{1}{|r|}{\textbf{Moderate IPC}} &
  28.7\% (2.39) &
  15.21\% (1.27) &
  \cellcolor[HTML]{B7D7A8}14.06\% (1.17) &
  \cellcolor[HTML]{E06666}42.03\% (3.5) \\ \hline
\multicolumn{1}{|r|}{\textbf{Frequent IPC}} &
  24.38\% (2.13) &
  \cellcolor[HTML]{E06666}33.73\% (2.95) &
  21.83\% (1.91) &
  \cellcolor[HTML]{B7D7A8}20.06\% (1.75) \\ \hline
\multicolumn{1}{|r|}{\textbf{Intense IPC}} &
  \cellcolor[HTML]{B7D7A8}8.06\% (0.6) &
  34.43\% (2.58) &
  \cellcolor[HTML]{E06666}34.54\% (2.59) &
  22.96\% (1.72) \\ \hline
\end{tabular}
}
\vspace{-15pt}
\end{table}

\pagebreak\noindent{\textbf{Code Snippet 4 - String Analysis}}

\snippetResults{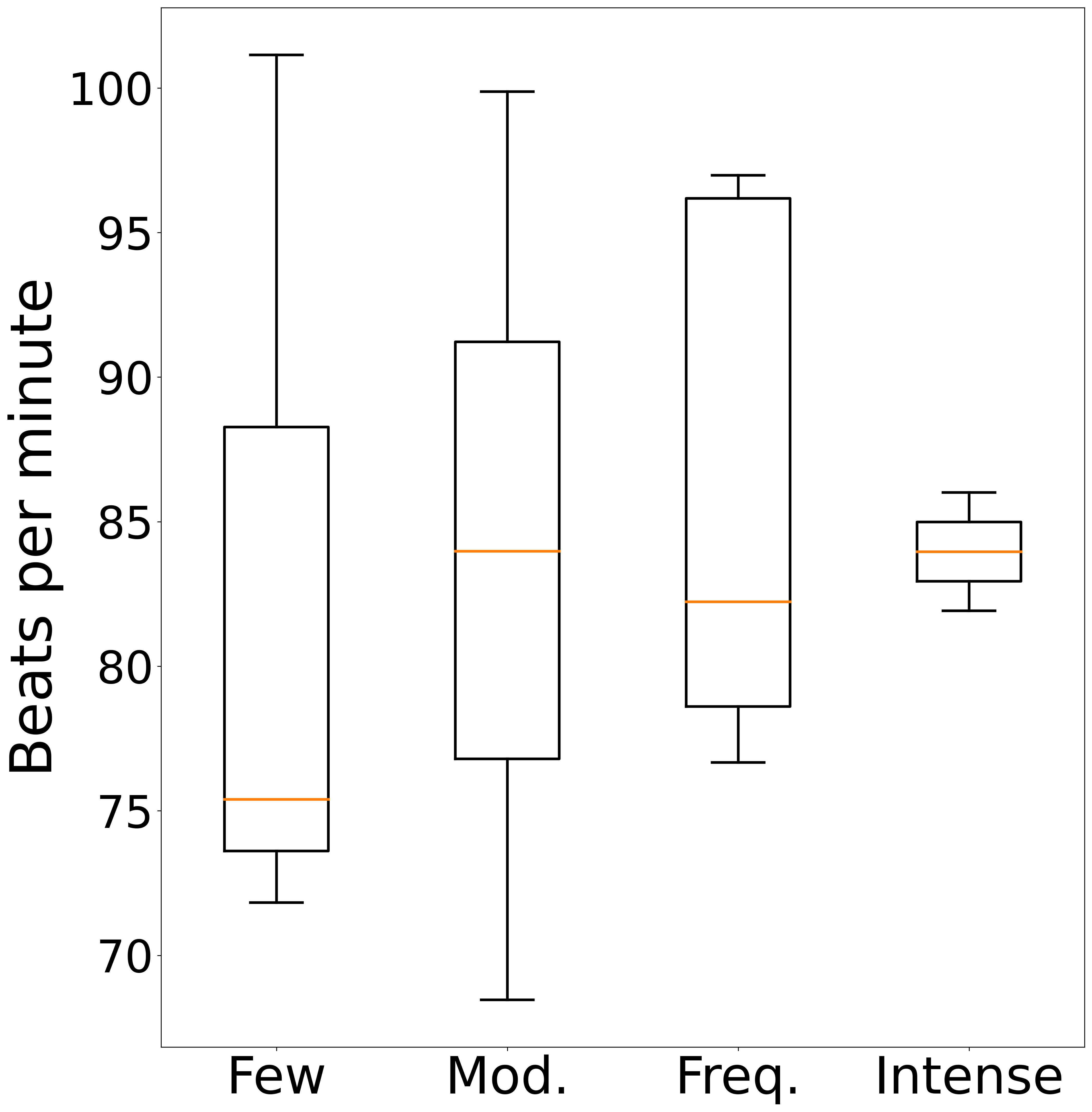}
Referring to Figure \ref{fig:biometrics_other_per_snippet_4}, seven participants answered this question correctly, including all three from Few IPC, three between Moderate and Frequent IPC, and one from Intense IPC. The median average heart rate was similar between Moderate, Frequent, and Intense IPC, but those in the Few IPC group had the lowest average heart rate. For time spent on the snippet, participants with Intense IPC had the longest average duration; participants with Few IPC had the shortest average duration.

As seen in Table \ref{tab:Snippet4_aois}, participants with Few, Moderate, and Frequent IPC spent the longest duration on variable assignments.  For code categories with the shortest duration, there was no consistency among the four IPC groups. 
\begin{table}
\vspace{-15pt}
\caption{Duration spent on a Code Category in Snippet 4}
\label{tab:Snippet4_aois}
\renewcommand{\arraystretch}{1.5} 
\resizebox{\columnwidth}{!}{
\scriptsize
\begin{tabular}{r|*{10}{>{\centering\arraybackslash}m{1.5cm}|}} 
\cline{2-10}
\multicolumn{1}{l|}{} &
  \textbf{Else Stmt} &
  \textbf{For Loop} &
  \textbf{If Stmt} &
  \textbf{Method Call} &
  \textbf{Method Decl} &
  \textbf{Return Stmt} &
  \textbf{Var Asgmt} &
  \textbf{Var Decl} &
  \textbf{Multiple} \\ \hline
\multicolumn{1}{|r|}{\textbf{Few IPC}} &
  \cellcolor[HTML]{B7D7A8}3.16\% (0.26) &
  7.95\% (0.65) &
  8.74\% (0.72) &
  9.61\% (0.79) &
  19.49\% (1.6) &
  17.82\% (1.46) &
  \cellcolor[HTML]{E06666}23.97\% (1.97) &
  7.7\% (0.63) &
  1.56\% (0.13) \\ \hline
\multicolumn{1}{|r|}{\textbf{Moderate IPC}} &
  11.28\% (1.83) &
  10.19\% (1.66) &
  9.96\% (1.62) &
  11.15\% (1.81) &
  11.59\% (1.88) &
  9.78\% (1.59) &
  \cellcolor[HTML]{E06666}20.56\% (3.34) &
  \cellcolor[HTML]{B7D7A8}9.7\% (1.58) &
  5.76\% (0.94) \\ \hline
\multicolumn{1}{|r|}{\textbf{Frequent IPC}} &
  14.72\% (2.29) &
  \cellcolor[HTML]{B7D7A8}4.6\% (0.72) &
  9.42\% (1.47) &
  9\% (1.4) &
  12.65\% (1.97) &
  10.71\% (1.67) &
  \cellcolor[HTML]{E06666}25.78\% (4.02) &
  10.58\% (1.65) &
  2.54\% (0.4) \\ \hline
\multicolumn{1}{|r|}{\textbf{Intense IPC}} &
  15.12\% (1.52) &
  7.69\% (0.78) &
  \cellcolor[HTML]{B7D7A8}5.32\% (0.54) &
  7.72\% (0.78) &
  \cellcolor[HTML]{E06666}24.46\% (2.47) &
  8.23\% (0.83) &
  12.54\% (1.26) &
  15.35\% (1.55) &
  3.57\% (0.36) \\ \hline
\end{tabular}
}\vspace{-15pt}
\end{table}

\pagebreak\noindent{\textbf{Code Snippet 5 - Palindrome}}

\snippetResults{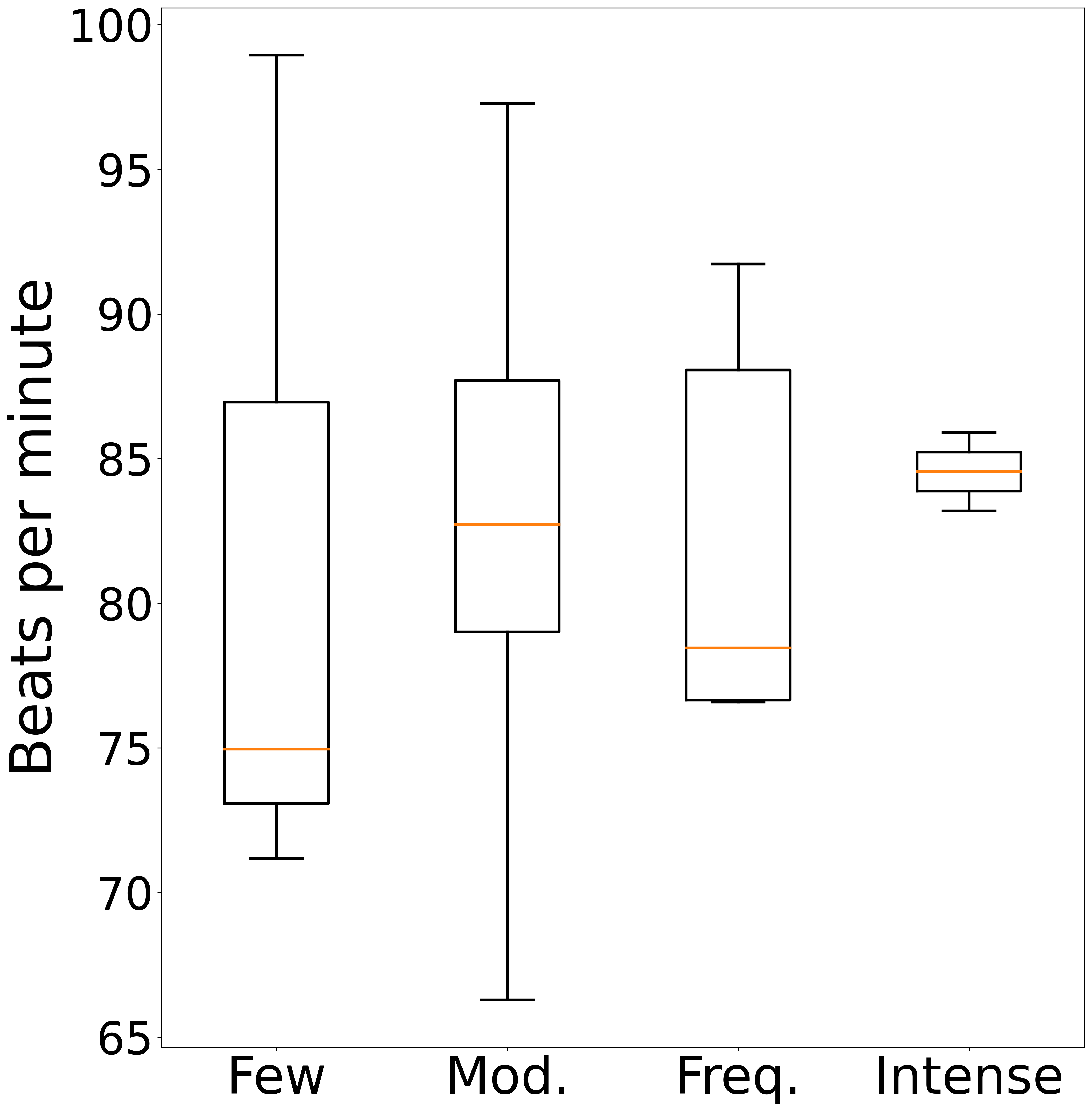}
For Code Snippet 5, eight participants answered correctly, again including all three from Few IPC, one from Moderate IPC, and four from Frequent IPC (See Figure \ref{fig:biometrics_other_per_snippet_5}). No participant from the intense IPC group answered correctly. Those in the few IPC group had the lowest average heart rate, whereas those in the intense IPC had the highest. Similar to the other snippets, participants with Few IPC spent the least amount of time on the snippet, whereas those with Intense IPC spent the most amount of time.

Referring to Table \ref{tab:Snippet5_aois} all four IPC groupings spent the most time looking at method declarations. Those with Few and Moderate IPC spent the least amount of time on return statements, whereas those with Frequent and Intense IPC spent the least amount of time on if statements. 
\begin{table}
\vspace{-15pt}
\caption{Duration spent on a Code Category in Snippet 5}
\label{tab:Snippet5_aois}
\renewcommand{\arraystretch}{1.2} 
\resizebox{\columnwidth}{!}{
\scriptsize
\begin{tabular}{r|*{8}{>{\centering\arraybackslash}m{1.5cm}|}}
\cline{2-8}
\multicolumn{1}{l|}{} &
  \textbf{Else Stmt} &
  \textbf{For Loop} &
  \textbf{If Stmt} &
  \textbf{Method Decl} &
  \textbf{Return Stmt} &
  \textbf{Var Asgmt} &
  \textbf{Var Decl} \\ \hline
\multicolumn{1}{|r|}{\textbf{Few IPC}} &
  \cellcolor[HTML]{B7D7A8}\vspace{3pt}5.5\% (0.33)\vspace{3pt} &
  \vspace{3pt}12.38\% (0.75)\vspace{3pt} &
  \vspace{3pt}6.02\% (0.37)\vspace{3pt} &
  \cellcolor[HTML]{E06666}\vspace{3pt}44.68\% (2.72)\vspace{3pt} &
  \cellcolor[HTML]{B7D7A8}\vspace{3pt}5.4\% (0.33)\vspace{3pt} &
  \vspace{3pt}7.73\% (0.47)\vspace{3pt} &
  \vspace{3pt}18.28\% (1.11)\vspace{3pt} \\ \hline
\multicolumn{1}{|r|}{\textbf{Moderate IPC}} &
  \vspace{3pt}14.16\% (1.81)\vspace{3pt} &
  \vspace{3pt}10.32\% (1.32)\vspace{3pt} &
  \vspace{3pt}11.6\% (1.48)\vspace{3pt} &
  \cellcolor[HTML]{E06666}\vspace{3pt}25.57\% (3.26)\vspace{3pt} &
  \cellcolor[HTML]{B7D7A8}\vspace{3pt}9.03\% (1.15)\vspace{3pt} &
  \vspace{3pt}14.81\% (1.89)\vspace{3pt} &
  \vspace{3pt}14.51\% (1.85)\vspace{3pt} \\ \hline
\multicolumn{1}{|r|}{\textbf{Frequent IPC}} &
  \vspace{3pt}10.19\% (0.74)\vspace{3pt} &
  \cellcolor[HTML]{B7D7A8}\vspace{3pt}6.34\% (0.46)\vspace{3pt} &
  \cellcolor[HTML]{B7D7A8}\vspace{3pt}6.34\% (0.46)\vspace{3pt} &
  \cellcolor[HTML]{E06666}\vspace{3pt}37.46\% (2.74)\vspace{3pt} &
  \vspace{3pt}7.27\% (0.53)\vspace{3pt} &
  \vspace{3pt}8.99\% (0.66)\vspace{3pt} &
  \vspace{3pt}23.41\% (1.71)\vspace{3pt} \\ \hline
\multicolumn{1}{|r|}{\textbf{Intense IPC}} &
  \vspace{3pt}20.88\% (2.89)\vspace{3pt} &
  \vspace{3pt}3.94\% (0.54)\vspace{3pt} &
  \cellcolor[HTML]{B7D7A8}\vspace{3pt}1.67\% (0.23)\vspace{3pt} &
  \cellcolor[HTML]{E06666}\vspace{3pt}40.36\% (5.58)\vspace{3pt} &
  \vspace{3pt}9.97\% (1.38)\vspace{3pt} &
  \vspace{3pt}7.14\% (0.99)\vspace{3pt} &
  \vspace{3pt}16.04\% (2.22)\vspace{3pt} \\ \hline
\end{tabular}
}\vspace{-15pt}
\end{table}

\subsubsection{\textbf{Overall Analysis.}}
\begin{figure}[H]
        \setlength{\abovecaptionskip}{0pt} 
        \setlength{\belowcaptionskip}{0pt} 
         \centering    
         \begin{subfigure}[b]{0.3\textwidth}
             \centering
             \includegraphics[width=\textwidth]{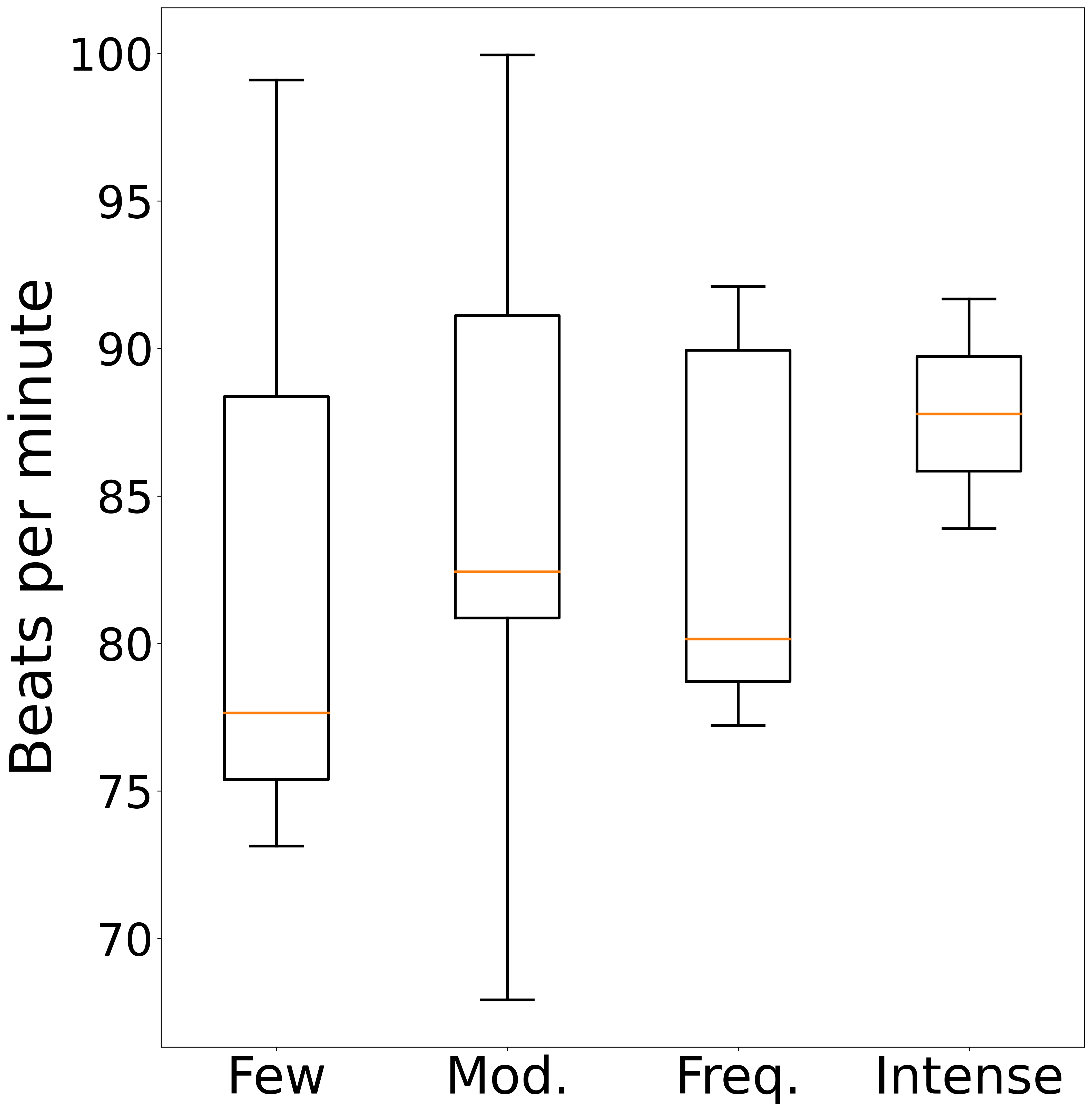}
             \caption{$Avg Heart Rate$}
             \label{fig:Average Heart rate}
         \end{subfigure}
         \hfill
         \begin{subfigure}[b]{0.3\textwidth}
             \centering
             \includegraphics[width=\textwidth]{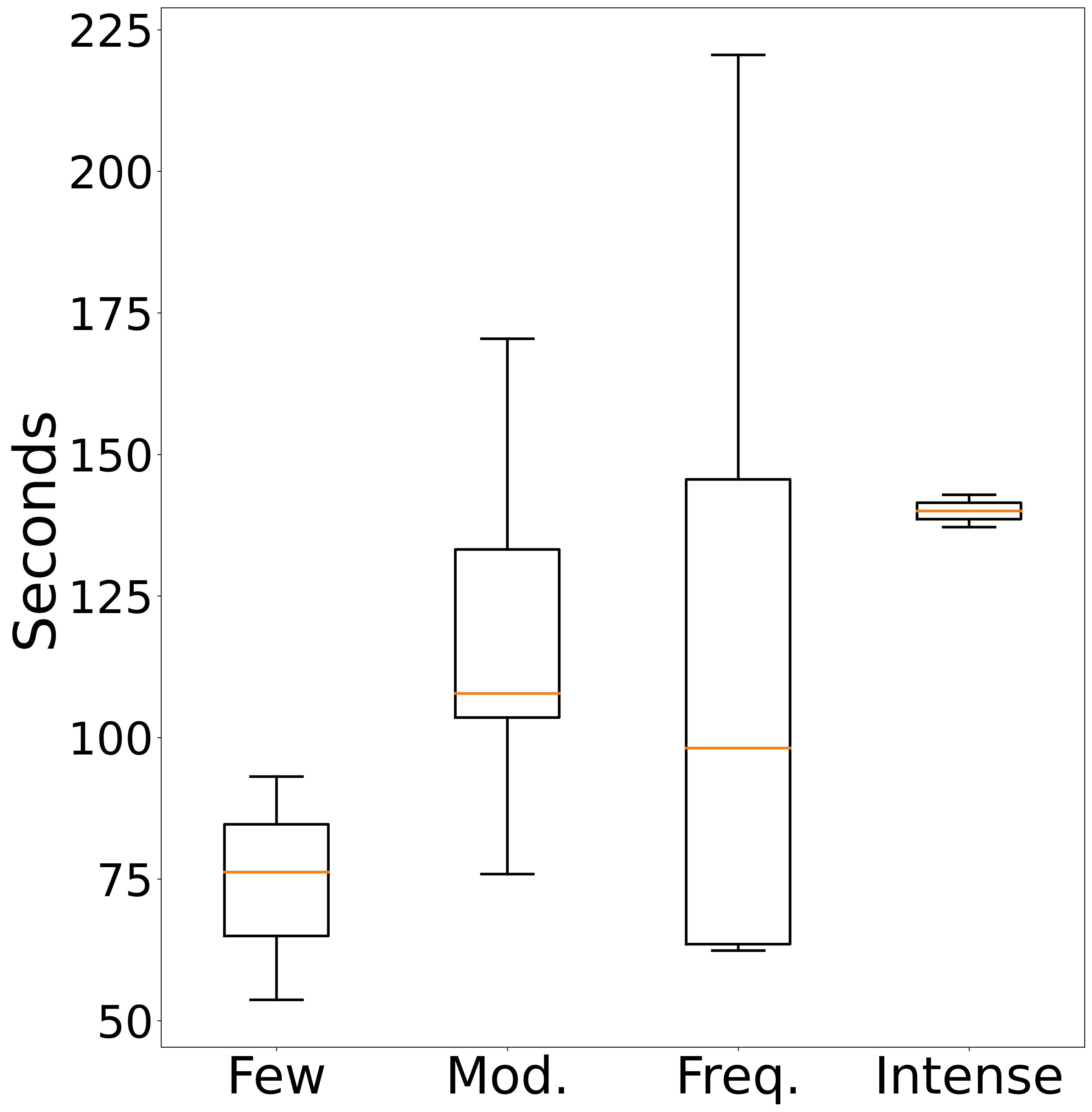}
             \caption{$Avg Time Spent$}
             \label{fig:Average Time Spent}
         \end{subfigure}
        \hfill
         \begin{subfigure}[b]{0.3\textwidth}
             \centering
             \includegraphics[width=\textwidth]{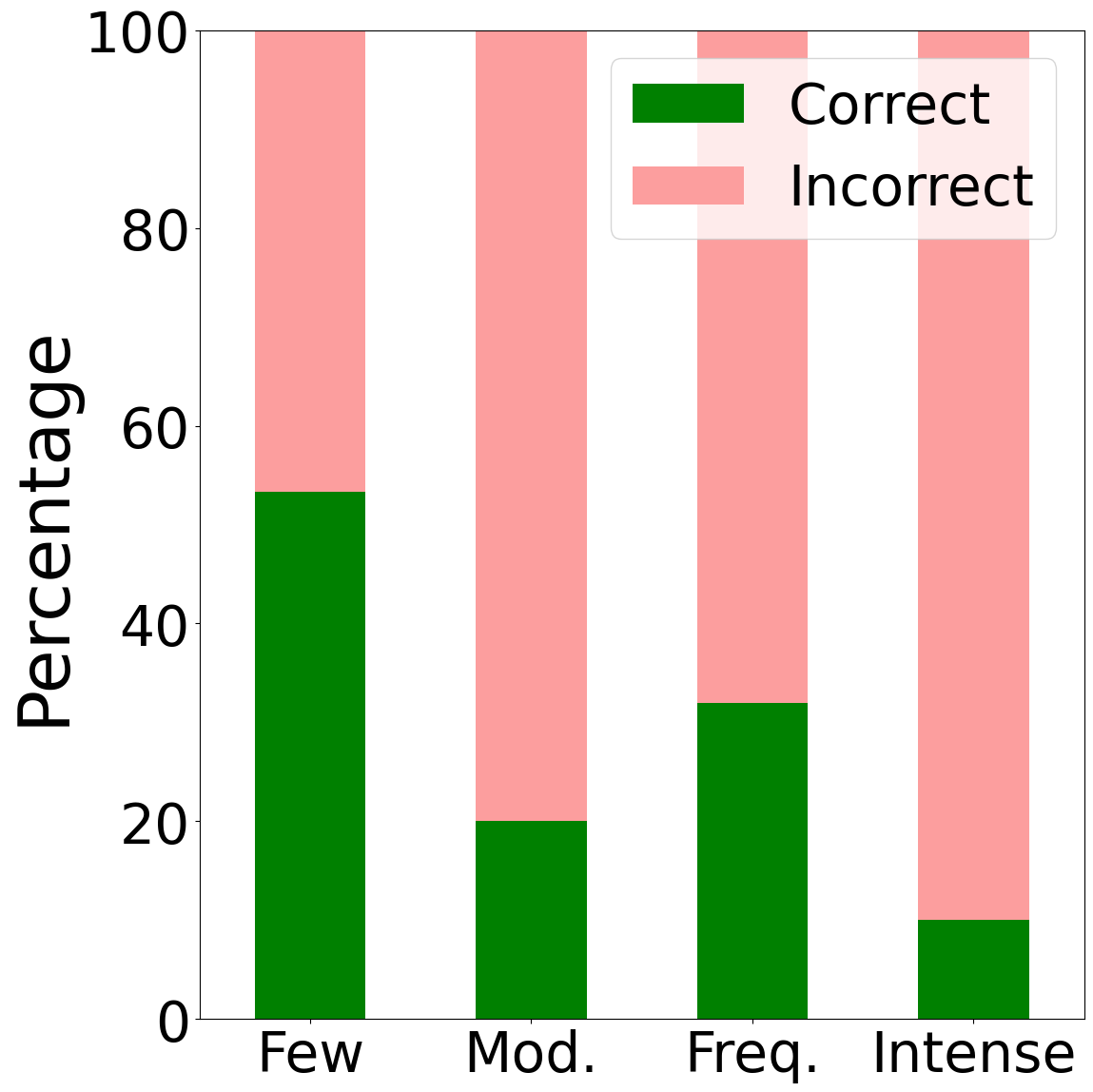}
             \caption{$Correctly Answered$}
             \label{fig: Correctly Answered}
         \end{subfigure} 
        
            \vspace{0.5cm} 
         
         \begin{subfigure}[b]{1\textwidth}
             \centering
             \includegraphics[width=\textwidth]{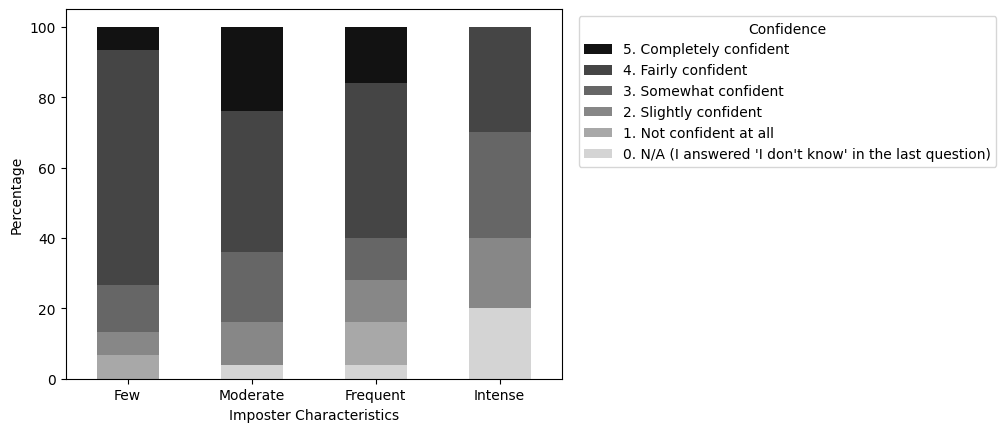}
             \caption{$Confidence in All Snippets$}
             \label{fig: Confidence in All Snippets}
         \end{subfigure} 
            \caption{Combined Results}
            \label{fig:Combined Results}
    \end{figure}

When averaging time across all 5 code snippets and for each imposter syndrome level, there appears to be a trend where the median of the time spent increases as participants score higher on Clance IP Scale (Figure \ref{fig:Average Time Spent}). When averaging how many times participants in each IPC group answered questions correctly across all code snippets, there was a downward trend (Figure \ref{fig: Correctly Answered}). To measure the extent of the relationship between these variables, we conducted a Pearson correlation test. The Pearson correlation test yielded statistically significant (i.e., p-value < 0.05) correlation coefficients of 0.52, equating to a moderate positive correlation between IPC score and time spent on a snippet and -0.52, equating to a moderate negative correlation between IPC score and average correctness. While in Figure \ref{fig:Average Heart rate}, there appears to be a positive trend between average heart rate and IPC, it was not statistically significant. Similarly, while there seems to be a pattern of those with high IPC being less confident in their answers in \ref{fig: Confidence in All Snippets}, no significant results were found.

To gain a deeper insight into overall how participants of varying IPC scores read code, a Pearson Correlation 2-tailed test was conducted to compare average normalized durations for each AOI code category (e.g., else-statement, for-loop). The test yielded a statistically significant (i.e., p-value < 0.05) correlation coefficient: 0.56 for the IPC score and Method Calls, equating to a positive moderate correlation, and 0.52 for the IPC score and Method Declarations. This could indicate that those with higher IPC scores tend to spend more time looking at code associated with these two categories.

\begin{tcolorbox}[top=0.5pt,bottom=0.5pt,left=1pt,right=1pt]
\textbf{Summary for RQ2.}
Students with higher characteristics of imposter syndrome were associated with spending longer on snippets and were less likely to answer the question related to each code snippet correctly. Across all code snippets, those with higher characteristics of imposter syndrome were more likely to look at the Method Call and Method Declaration code categories.
\end{tcolorbox}

\section{Threats To Validity}
\label{Section:threats}

Although our study sample size is small and limited to a single institute, it remains valuable as an initial exploratory study providing a foundation for further research. Furthermore, research shows that program comprehension studies using an eye tracker have an average of 18.08 participants with a standard deviation of 9.90 \cite{Obaidellah2019}. Likewise, due to the small sample size, the small gender and race/ethnicity distribution in our participants' pool may not be representative of the general population. The number and type of code snippets used in this study in order to evaluate comprehension threaten the study's validity, as they may not represent real-world codebases that students will encounter in their careers. However, these code snippets represent basic programming concepts that students learn when preparing for their degree and are commonly asked in entry-level software engineering industry interviews.

The validity of our study may also be threatened by the limitations of the Gazepoint GP3 HD Eye Tracker and Biometrics Kit. However, these are research-grade devices that have been previously utilized in scientific publications \cite{Gazepoint,Gazepoint2}. Overall, while our study has some inherent limitations as an initial investigation, it establishes a valuable foundation for further research on this topic. Finally, as the experiment was conducted in a controlled lab environment, it may not fully reflect real-world software development situations and, therefore, could impact the validity of our results. However, the controlled setting helped us focus on the effects of imposter syndrome on code comprehension. Moreover, we followed established guidelines to set up our eye-tracking experiment \cite{sharafi2020practical}.

\section{Conclusion \& Future Work}
\label{Section:conclusion}
Imposter syndrome is a psychological barrier that can negatively affect the performance of students and professionals. While research on imposter syndrome in software engineering does exist, little is known about how it affects code comprehension cognition. In this exploratory study, we examine the level of imposter syndrome final-year undergraduate computer science students exhibit when comprehending code. We further measure the cognitive impact using an eye tracker and heart rate monitor. Our findings show that: (1) students identifying as males show lower levels of imposter syndrome, (2) higher levels of imposter syndrome are associated with increased duration of time spent on a snippet and lower chances of solving the problem correctly, and (3) those with higher imposter syndrome levels are more likely to look at method declarations and method calls. While limited in scope, our study establishes a foundation for further research on imposter syndrome and its effects on core software engineering competencies.

Our future work involves working with industry professionals in a similar study to understand how imposter syndrome affects computer science professionals in their day-to-day work, including its impact on mental health.

\section{Acknowledgments}
\label{Section:acknowledgments}
We thank students who took the time to participate in our study.

\section{Data Availability Statements}
\label{Section:das}

The high-resolution images of the code snippets, survey questionnaires, and participant data are available at: \cite{ProjectWebsite}.

\bibliographystyle{splncs04}
\bibliography{main.bib}  

\end{document}